\documentclass[draft]{agujournal2019}
\usepackage{url} 
\usepackage{lineno}
\usepackage[inline]{trackchanges} 
\usepackage{soul}
\usepackage{enumitem}
\draftfalse

\journalname{Earth and Space Science}

\begin{document}
\justify
%
%


\title{Quasi-constant time gap in multiple rings of elves}

%
%




\authors{The Pierre Auger Collaboration}
\affiliation{1}{The Pierre Auger Observatory, Av. San Martín Norte 306, 5613 Malargüe, Mendoza, Argentina; http://www.auger.org}





\correspondingauthor{The Pierre Auger Collaboration}{spokespersons@auger.org}



\begin{keypoints}
\item Exploiting a 100 ns time resolution multi-elves are shown to be very frequent, approximately 23\% of the total.
\item Observations of multi-elves in Argentina challenge the model attributing their origin to reflection of electromagnetic pulses at the ground.
\item The multiple peaks observed in elves appear to be more closely linked to the lightning waveform than to its source altitude.
\end{keypoints}

%
%

%
%


\begin{abstract}
We present evidence that the time delay between the multiple rings of elves is not caused by the ground reflection of the electromagnetic pulse produced by intracloud lightning. To investigate temporal differences of multi-elves, we analyzed data from four storms occurring at various times and distances from the Pierre Auger Observatory in Malargüe, Argentina. The Auger fluorescence detector's high temporal resolution of 100 ns enabled the frequent observation of multi-elves, accounting for approximately 23\% of the events. By examining the traces of 70 double and 24 triple elves, we demonstrate that the time delay between the rings remains relatively constant regardless of the arc distance to the lightning. These results deviate from the trend expected from the electromagnetic pulse (EMP) ground reflection model, which predicts a decreasing time delay with increasing arc distance from an intracloud lightning at a given height. The first emission ring is due to a direct path of the EMP to the ionosphere, with the reflected EMP creating the second ring. Simulations conducted with this model demonstrate that short energetic in-cloud pulses can generate four-peak elves, and a temporal resolution of at least 25 $\mu$s is required to separate them. Therefore, temporal resolution is crucial in the study of multi-elves. Our observations in the Córdoba province, central Argentina, indicate that the current understanding of the mechanism generating these phenomena may be incomplete, and further studies are needed to assess whether multi-elves are more likely related to the waveform shape of the lightning than to its altitude.
\end{abstract}

\section{Introduction}\label{Introduction}
ELVES (Emission of Light and Very Low-Frequency perturbations due to Electromagnetic Pulse Sources) is a type of transient luminous event (TLE- elves, sprites, halos, blue jets), occurring when the electromagnetic pulse (EMP) from an intense lightning strike reaches the lower ionosphere, causing the ionospheric electrons to heat up and produce secondary ionization and optical emissions~\cite{inan1991heating, TaranenkoEtal1993, FukunishiEtal1996}. Evidence shows that elves mainly originated from negative cloud-to-ground (CG) lightning~\cite{barrington1999elves, Newsome2010}, as opposed to sprites which are usually triggered by positive CG. Elves are the most common type of TLE, occurring globally at a rate of 35 elves/min, while sprites and halos occur at 1 event/min~\cite{chen2008}.

The elves emission pulses usually last about 1 ms, and several authors~\cite{barrington1999elves, Newsome2010, Lyu2015, Merenda2020A} have reported cases where two of the pulses appear close together in time, sometimes only tens or a few hundred microseconds apart. These pulses, often called elve doublets, are thought to be associated with the same event because they are too close to be caused by multiple return strokes since the interstroke period is around 60 ms, and to be observed, detectors need resolutions better than a few tens of microseconds. 

The origin of the multiple elves phenomenon remains under investigation. \citeA{Marshall2015} suggested that multiple elves may be generated by compact intracloud lightning discharges (CIDs), with the time difference, $\Delta T$, between the rings of a doublet being directly related to the height of the source. Using the maximum value of the $\Delta T$ detected among the vertical and horizontal channels of the Photometric Imager of Precipitated Electron Radiation (PIPER)~\cite{MarshallEtal2008}, ~\citeA{Marshall2015} calculated the lightning height and confirmed observationally that $\Delta T$ correlates with high-altitude CIDs. From these observations, they also suggest that many simple elves labelled as CG-lightning generated are CIDs, but the temporal resolution of the PIPER (40 $\mu$s) is not sufficient to separate the peaks of these elves. 

\citeA{Liu2017} used this approach to simulate the emissions in the ionosphere produced by energetic intracloud pulses (EIPs) and linked to terrestrial gamma-ray flashes. Results from the simulations of EIPs producing elves doublets suggest that at least 50 $\mu$s time resolution is needed to differentiate the two peaks. They also show that very short EIPs (33.6 $\mu$s) can produce elves quadruplets, and the minimum time resolution to separate the peaks is about 25 $\mu$s. Therefore, time resolution is crucial for the study of multi-elves.  

On the other hand, \citeA{marshall2012improved} showed that CGs with a very long rise time of the current waveform could produce double elves. However, \citeA{Marshall2015} noted that those rise times are unrealistically long but the possibility of a fraction of double elves originating from CG discharges remains open. 

Since 2013, the fluorescence detector (FD) at the Pierre Auger Observatory~\cite{pierre2015pierre} in the Mendoza province-Argentina has been catching elves using a dedicated trigger~\cite{MussaCiaccio2012}. The FD is well-suited for elves detection due to the optimization of its telescope parameters for recording faint light in the 300-420 nm band~\cite{AbrahamEtal2010}. The elves' observation footprint covers an area of $3\times 10^6$ km$^2$~\cite{Merenda2020A}, including the C\'ordoba province where large thunderstorms often occur~\cite{witze2018inside}. 

Furthermore, the FD telescopes offer an exceptional time resolution of 100 ns, surpassing other TLE detection systems such as the Imager of Sprites and Upper Atmospheric Lightning onboard the FORMOSAT-2 satellite (ISUAL), which has a 200 $\mu$s time resolution~\cite{chern2003global}, PIPER with 40 $\mu$s~\cite{MarshallEtal2008}, the Multiwavelength Imaging New Instrument for the Extreme Universe Space Observatory (Mini-EUSO) with 2.5 $\mu$s~\cite{capel2018mini}, and the Atmosphere-Space Interactions Monitor (ASIM) with 10 $\mu$s~\cite{neubertEtal2020}. This feature enables unprecedented precision in capturing the internal structure of elves. The Auger FD has successfully identified not only two-ring elves but also three-ring events, as evidenced by the first reported elve with three peaks in its traces~\cite{Merenda2020A}.

In this paper, we analyzed the traces of double and triple elves detected at the Auger Observatory, finding that the time delay between the peaks is relatively constant with the arc distance to the lightning source. This behaviour persists in the multi-elves of four storms at different distances from the observatory. These findings do not fit the proposed model in reference \citeA{Marshall2015} that predicts the time delay should decrease as the arc distance to the source increases, leading us to question the EMP reflection model as an explanation for the origin of multiple elves.

\section{Multi-Elves Data from the Auger Observatory}
The Pierre Auger Observatory is a ground-based cosmic-ray observatory that measures the properties of the most energetic particles in the universe and aims to discover their sources. These high-energy particles interact with the Earth's atmosphere creating extended air showers. The observatory uses a surface array of 1,600 water-Cherenkov detectors and FDs to measure air showers over 3,000 km$^2$~\cite{AbrahamEtal2010}.

The FD~\cite{AbrahamEtal2010} comprises 24 telescopes at four sites, namely Los Leones (LL), Los Morados (LM), Loma Amarilla (LA) and Coihueco (CO). Each building accommodates six telescopes, whose fields of view are 30$^{\circ}$ in azimuth $\times$ 28$^{\circ}$ in elevation. The UV optical filters have a bandwidth of 300–420 nm, and the photomultiplier tubes of the cameras—440 per telescope—have a quantum efficiency of 30\% within this range, limiting the red and infrared light detection from all TLEs. The readout system includes a 28 $\mu$s pedestal and, in 2017, its time window was extended from 300 $\mu$s to 900 $\mu$s to fully observe the high light intensity region for most of the elves. Although the viewing distances range from 3 to 30 km for cosmic rays, due to its high sensitivity to light, the Auger FD has detected elves at 250-1,000 km (details in Figure 1 of~\cite{Merenda2020A}). The FD is operational during locally clear nights with low background light and clear local conditions, accumulating about 1,200 hours of on-time over 12 months, equivalent to almost 15\% duty cycle. In addition, the observatory employs lasers, lidars, and infrared cloud cameras to measure the optical transparency of the atmosphere over the array. 

An additional fluorescence detector, the High Elevation Auger Telescopes (HEAT) \cite{AbrahamEtal2010} consisting of three telescopes set at a 45-degree elevation, is also part of the observatory. While the primary objective of HEAT is to detect low-energy cosmic ray showers, it also enables the detection of elves. The elves detected in HEAT originate from storms closer to the observatory, within a range of approximately 50 to 150 km. In December 2020, we implemented the elves detection trigger in HEAT, resulting in fascinating nearby events available for analysis.

As reported by~\citeA{Merenda2020A}, during 2013-16, most of the events detected in the Auger FD occurred during the southern summer, with production showing minimal variation between June and August. We identified April 27-28, 2020, as the night with the highest number of multi-elves between 2014 and 2020. Accordingly, we selected these dates along with two other thunderstorms from November 10, 2018, and December 20, 2019. Additionally, we included the events detected by HEAT on March 14, 2021, to examine multi-elves occurring closer to the observatory.

The total number of events from these four storms is 478, as shown in Table \ref{tab:data}, of which 369 are single and 109 are multi-elves. The occurrence of multi-elves over the total events (\%ME) is 23\%, consisting of 70 double, 24 triple elves, and 15 multi-elves occurring with halos labelled as 'Halos'. These events can be detected simultaneously within the 900 $\mu$s time window of the FD readout system. To avoid introducing biases in the analysis of the time delay between multi-elves peaks, we handle them separately, focusing on the multi-elves traces and disregarding the associated halo emissions. A new dedicated camera is currently capturing such events, which will be analyzed in detail in forthcoming publications.  

\begin{table}[h]
\caption{Total numbers of single, double, and triple elves detected across four storms at various dates and distances from the Auger FD are reported. Multi-elves with halos are denoted as 'Halos'. The percentage of multi-elves relative to the total (\% ME) varies for each storm. Overall, multi-elves account for 23\% of the total detected across the four storms.}
\label{tab:data}
\centering
\begin{tabular}{lcccccc}
\textbf{Storm date}  & \textbf{Total} & \textbf{Singles} & \textbf{Doubles} & \textbf{Triples} &\textbf{Halos} & \textbf{\% ME} \\ \hline
November 10, 2018    & 114 & 96 & 11 & 3 &4 & 16\%\\ 
(01:26:46 - 07:49:03) UTC  &  &  &  & & \\ \hline
December 20, 2019    & 156 & 134 & 16 & 3& 3 & 14\%\\ 
(01:56:40 - 06:14:18) UTC  &  &  &  & & \\ \hline
April 27-28, 2020    & 137 & 77 & 35 & 17 & 8& 44\%\\ 
(23:36:58 - 09:36:05) UTC  &  &  &  & & \\ \hline
March 14, 2021       & 71 & 62 & 8 & 1 & 0 & 13\% \\ 
(00:51:03 - 06:34:07) UTC  &  &  &  & & & \\ \hline
\textbf{Total events 
analyzed} &  \textbf{478} &\textbf{369}  & \textbf{70} & \textbf{24} &\textbf{15}& \textbf{23\%}\\ 
\end{tabular}
\end{table}

The number of single and multi-elves within different storms display notable variability. In the November and December storms, \%ME is about 16\% and 14\%, respectively, whereas, in April, this proportion significantly increases to 44\%. Within the HEAT observations conducted in March, the percentage of multi-elves is 13\%. These findings highlight the prevalence of multi-elves phenomena within the analyzed data, indicating that they often appear during the storms that the Auger Observatory can investigate.

\section{Multi-Elves Traces Analysis}\label{Algorithm}
\begin{figure}[h]
    \centering
    \includegraphics[scale=0.35]{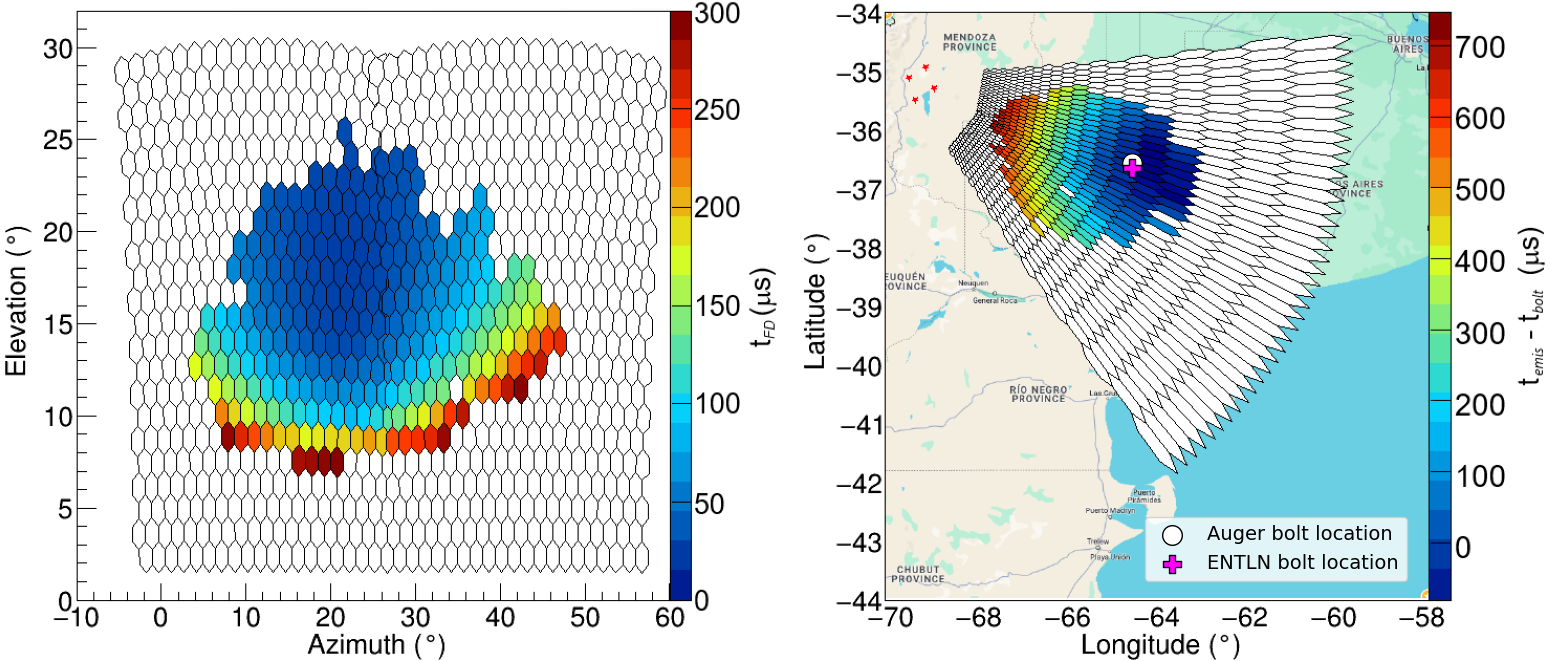}
    \caption{Left panel: pixels of two FD cameras triggered by a typical single elve. This event was detected on April 28, 2020, at 04:08:13 UTC. The colours show the time evolution of the light as seen in the FD cameras, which usually lasts around 300 $\mu$s. Right panel: time evolution of the event as projected at the base of the ionosphere fixed at an altitude of 92 km. After correcting for the transit time from the emission layer to the FD, we observe the concentric spread of light from the source lightning at 36.57$^\circ$ S and 64.54$^\circ$ W (Auger bolt location). The Earth Networks (ENTLN) lightning location correlated with this event, 36.64$^\circ$ S and 64.53$^\circ$ W, aligns with our determined location. The red stars indicate the locations of the four FD buildings.}
    \label{fig:elves}
\end{figure}

The Auger FD employs a multi-level trigger system for event selection, and elves are primarily detected by the third-level trigger originally developed as a lightning veto. The algorithm searches for events with a radially expanding light front. It identifies the first triggered pixel in the camera and requires adjacent pixels' pulse start times to show monotonic growth. At least three neighbouring pixels on both sides of the first signal (or one side if near the camera edge) and three above and below must satisfy the described cut~\cite{Merenda2020A}. The left panel of Figure \ref{fig:elves} shows an example of pixels triggered by the light of a single elve on April 28, 2020, at 04:08:13 UTC in two FD cameras. This plot shows the time corresponding to the peak of the light pulse at each pixel ($t_{FD}$), with the typical time for elves around 300 $\mu$s.

Using the traces of elves, we can determine the location and timing of the lightning source from the first triggered pixel (details in reference~\cite{Merenda2020A}). The right panel in Figure \ref{fig:elves} shows the projection of the light of the single elve onto the ionosphere (at a height of 92 km) spreading concentrically outward from the source lightning located at 36.57$^\circ$ S latitude and 64.54$^\circ$ W longitude. We correlated this single elve event with the Earth Networks Total Lightning Network (ENTLN) data, and the plot shows that the reported location (36.64$^\circ$ S and 64.53$^\circ$ W) is consistent with the one we obtained (Auger bolt location).

\begin{figure}
    \centering
    \includegraphics[scale=0.5]{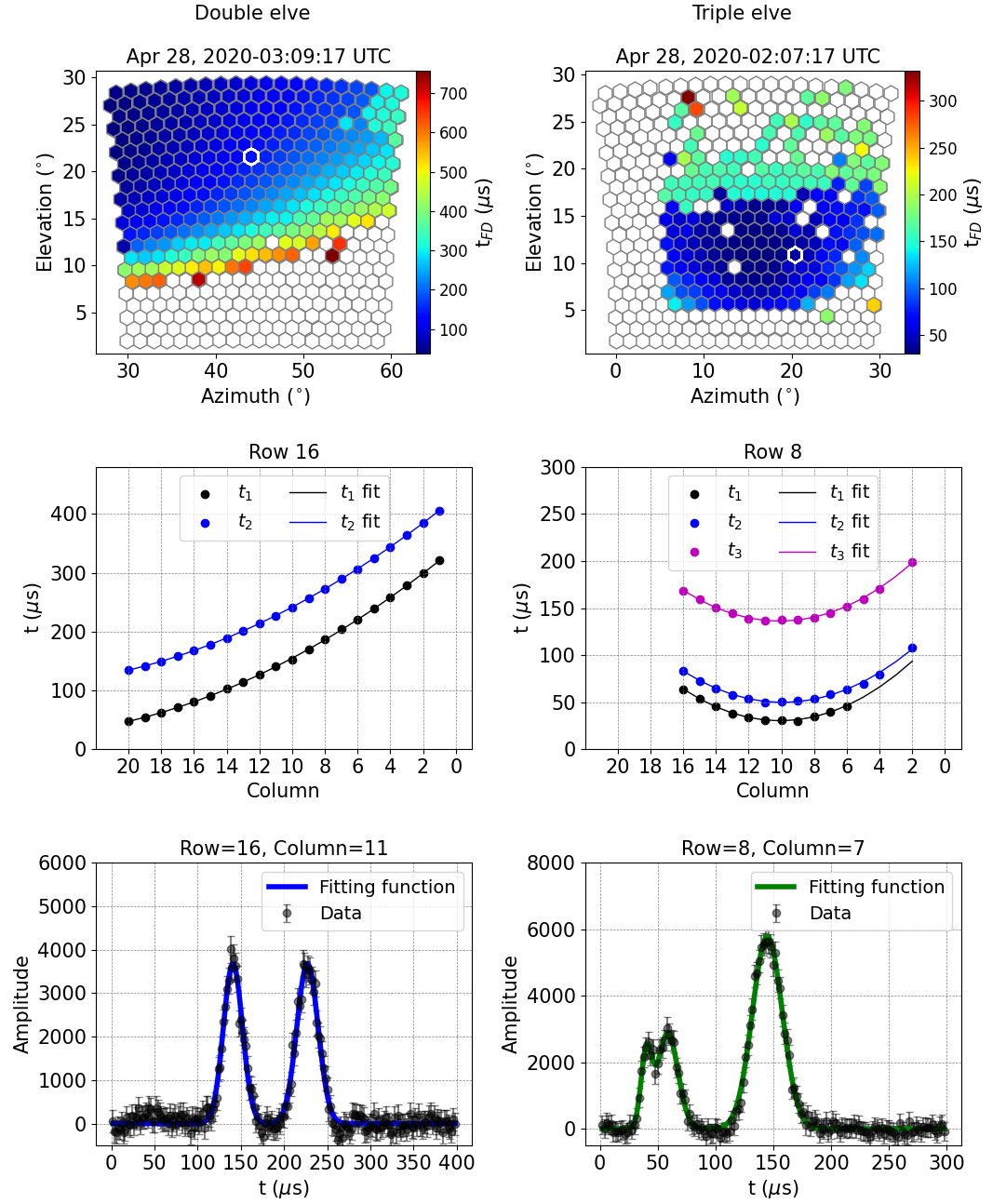}
    \caption{Top panel: time evolution of the first peak of a double elve event detected by an FD telescope on April 28, 2020, at 03:09:17 UTC (left), and a triple elve event at 02:07:17 UTC (right). The highlighted pixel in each camera was selected to illustrate typical double-peak and triple-peak signals. Middle panel: examples of prominent peaks $t_1$ and $t_2$ for the double elve in row 16 pixels, and $t_1$, $t_2$, $t_3$ for the triple elve in row 8 pixels. Row numbers increase with pixel elevation, and column numbers decrease with azimuth, as shown in the top panel. The distinct parabolic patterns of the elves emission are visible in both cases ($t_1$ fit, $t_2$ fit and $t_3$ fit). Bottom panel: examples of Gaussian fitting for the double elve signal in row 16, column 11, and the triple elve signal in row 8, column 7.}
    \label{fig:peakfinder}
\end{figure}

With the lightning location determined, we can analyze the time differences between the multiple rings of an event and its source. This process begins with identifying the most prominent peaks in the elve traces, labelled as $t_1$, $t_2$, and $t_3$. Figure \ref{fig:peakfinder} illustrates in the top panel the temporal evolution of the first light pulse in a double elve (left) and a triple elve (right). The camera pixels are indexed according to the row number (from 1 to 22), which increases from bottom to top in elevation, and the column number (from 1 to 20), which increases from right to left in azimuth. We selected row 16 of the double-peak event occurring on April 28, 2020, at 03:09:17 UTC, to illustrate in the middle panel examples of determining $t_1$ and $t_2$. For the triple-peak event occurring at 02:07:17 UTC, we show $t_1$, $t_2$, and $t_3$ in row 8. In both cases, the characteristic parabolic shapes of the elves are evident ($t_1$ fit, $t_2$ fit and $t_3$ fit).

It is important to note that, not all pixels in a multi-elve event will exhibit the distinct peaks required for the algorithm to differentiate them based on their prominence and width. For example, in the triple elve event in row 8, columns 4 and 5, $t_1$ is not distinguishable, whereas $t_2$ and $t_3$ are identifiable. This issue is not attributable to the inefficiency of our algorithm, which has demonstrated effectiveness in resolving separate peaks with a precision of approximately 7 $\mu$s. Instead, this limitation is primarily due to the relatively low prominence of $t_1$ compared to $t_2$ and $t_3$ in some pixels.

After finding the peaks, we perform the corresponding Gaussian fitting (single, double, or triple) using $t_1$, $t_2$ and $t_3$ as the initial parameters. We select the fit that minimizes the error relative to the signal data, from which we obtain the values for the times ($t_1$, $t_2$, and $t_3$), amplitudes ($a_1$, $a_2$, and $a_3$), and peak widths ($\sigma_1$, $\sigma_2$, and $\sigma_3$) of the pulses. The bottom panel in Figure \ref{fig:peakfinder} provides examples of Gaussian functions fitted to the data for one pixel of each event. On the left, a double Gaussian fit corresponds to the signal of the double elve in column 11 and row 16, while on the right, a triple Gaussian fit corresponds to the signal of the triple elve in column 7 and row 8. The amplitude of these signals is expressed in the number of photons.

Finally, we calculate the time difference between the peaks of double elves ($\Delta T_d = t_2 - t_1$), and for triple elves, we determine both $\Delta T_d$ and $\Delta T_t = t_3 - t_1$.

\section{Time Delay of Double and Triple Elves}
To analyze the behaviour of the time delay of the multi-elves with the distance to the lightning source, we recall the model proposed by~\citeA{Marshall2015} where the occurrence of two-peaked elves can be attributed to intracloud lightning strokes at a height denoted as $h_\mathrm{s}$. In this model, the first peak in the elve trace is created by the electromagnetic pulse taking a direct path to the ionosphere. The second peak is generated by the ground reflection of the EMP, which reaches the ionosphere with a time delay ($\Delta T$). 

\begin{equation}
\label{eq:distances}
d_{2,1}^{2}=\left(R_{\mathrm{E}} \pm h_\mathrm{s}\right)^{2}+\left(R_{\mathrm{E}}+h_{\mathrm{iono}}\right)^{2}-2\left(R_{\mathrm{E}} \pm h_\mathrm{s}\right)\left(R_{\mathrm{E}}+h_{\mathrm{iono}}\right) \cos\left(\frac{\mathrm{D_{arc}}}{R_{\mathrm{E}}+h_{\mathrm{iono}}}\right),
\end{equation}
\begin{equation}
\label{eq:dt}
  \Delta T = \frac{d_2-d_1}{c}.
\end{equation}

\begin{figure}[h]
    \centering
    \includegraphics[scale=0.5]{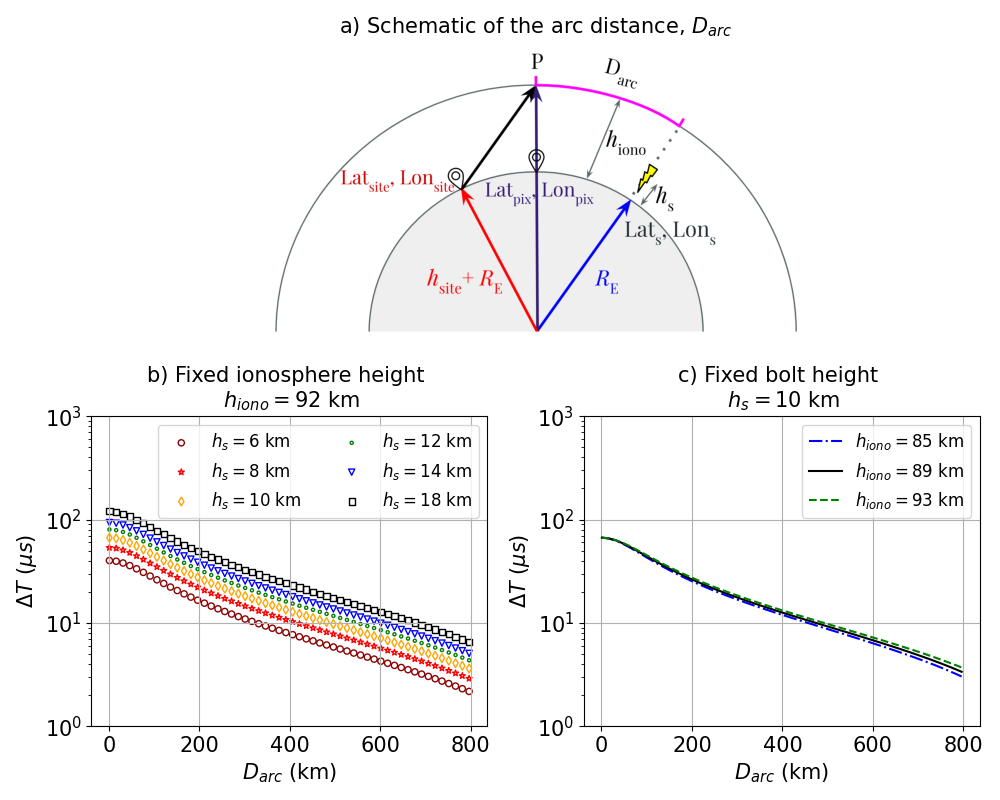}
    \caption{a) Schematic of the arc distance ($\mathrm{D_{arc}}$) between the lightning source ($\mathrm{Lat_s}$, $\mathrm{Lon_s}$) and the emission point at the ionosphere P. We determine the position of point P ($\mathrm{Lat_{pix}}$, $\mathrm{Lon_{pix}}$) by projecting a camera pixel's position onto the ionosphere, using the associated site coordinates ($\mathrm{Lat_{site}}$, $\mathrm{Lon_{site}}$). Additionally, we calculate the lightning location ($\mathrm{Lat_s}$, $\mathrm{Lon_s}$) based on the observed light-time distributions of recorded elves. b) Time delay between two peaks of an elve, given by the equations \ref{eq:distances} and \ref{eq:dt}, with the ionosphere height fixed at 92 km and various values of lightning height ($h_\mathrm{s}$). c) Varying the ionosphere height does not significantly change the functionality of the curves of $\Delta T$ with distance $\mathrm{D_{arc}}$.}
    \label{fig:bouncemechanism}
\end{figure}

Equations \ref{eq:distances} and \ref{eq:dt} describe the relationship between $\Delta T$ and $h_\mathrm{s}$. In these equations, $d_{2}$ and $d_{1}$ represent the distances travelled by the EMP in the direct-to-source and ground reflection paths, respectively. Additional parameters include $R_{\mathrm{E}}$, the Earth's radius; $h_{\mathrm{iono}}$, the height of the ionosphere; and $\mathrm{D_{arc}}$, the arc distance between the emission point P in the ionosphere and the lightning location, as illustrated schematically in panel a of Figure \ref{fig:bouncemechanism}. 

Each FD camera has 440 pixels that individually collect a part of the elve light ring. We obtain the position of point P ($\mathrm{Lat_{pix}}$, $\mathrm{Lon_{pix}}$) by projecting the position of a camera pixel onto the ionosphere, based on its elevation, azimuth and the coordinates of the site where the elve is detected
($\mathrm{Lat_{site}}$, $\mathrm{Lon_{site}}$). As mentioned earlier, we reconstruct the location of the lightning ($\mathrm{Lat_s}$, $\mathrm{Lon_s}$) from the measured light-time distributions of the recorded elve. 

Using equations \ref{eq:distances} and \ref{eq:dt}, we can determine how $\Delta T$ varies with the arc distance, $\mathrm{D_{arc}}$. For instance, a lightning source at a height of $h_\mathrm{s} = 6$~km can generate a double elve event with a maximum time delay of 40 $\mu$s, assuming the ionosphere is at a fixed height of 92~km. This behaviour is illustrated in panel b of Figure \ref{fig:bouncemechanism}, where the dependence of $\Delta T$ on $\mathrm{D_{arc}}$ is shown for various lightning heights ($h_\mathrm{s}$). Changes in ionosphere height do not affect the functional form of the curves describing the relationship between $\Delta T$ and $\mathrm{D_{arc}}$ distance, as shown in panel c.

This model has been used to explain the wide time separation between the peaks of double elves, observationally confirmed to correlate with high altitude compact intracloud lightning discharges~\cite{Marshall2015}. However, some authors report multi-elves with temporal differences that appear too large to be associated with realistic lightning heights. For example, \citeA{Newsome2010} reported a set of 40 double elves with $\Delta T$ between 80 $\mu$s and 160 $\mu$s detected by the PIPER with a time resolution of 40 $\mu$s. They argued that these events are unlikely to be caused due to a ground reflection mechanism, which implies an altitude of the source above 18~km, nor are they caused by multiple return strokes, which would be further separated in time.

Thanks to the temporal resolution of the FD, we report events over a wider range, with the smallest $\Delta T$ observed being 7 $\mu$s. The $\Delta T$ of a multi-elve must follow a behaviour given by the function in equation \ref{eq:distances} to consider that its origin is the EMP bounce mechanism. However, we have not found double elve events with this characteristic in the Auger data, since the $\Delta T_d$ tends to be constant with the arc distance $\mathrm{D_{arc}}$. We also observe this behaviour in the $\Delta T_d$ and $\Delta T_t$ values of triple elves. 

\begin{figure}
    \centering
    \includegraphics[scale=0.5]{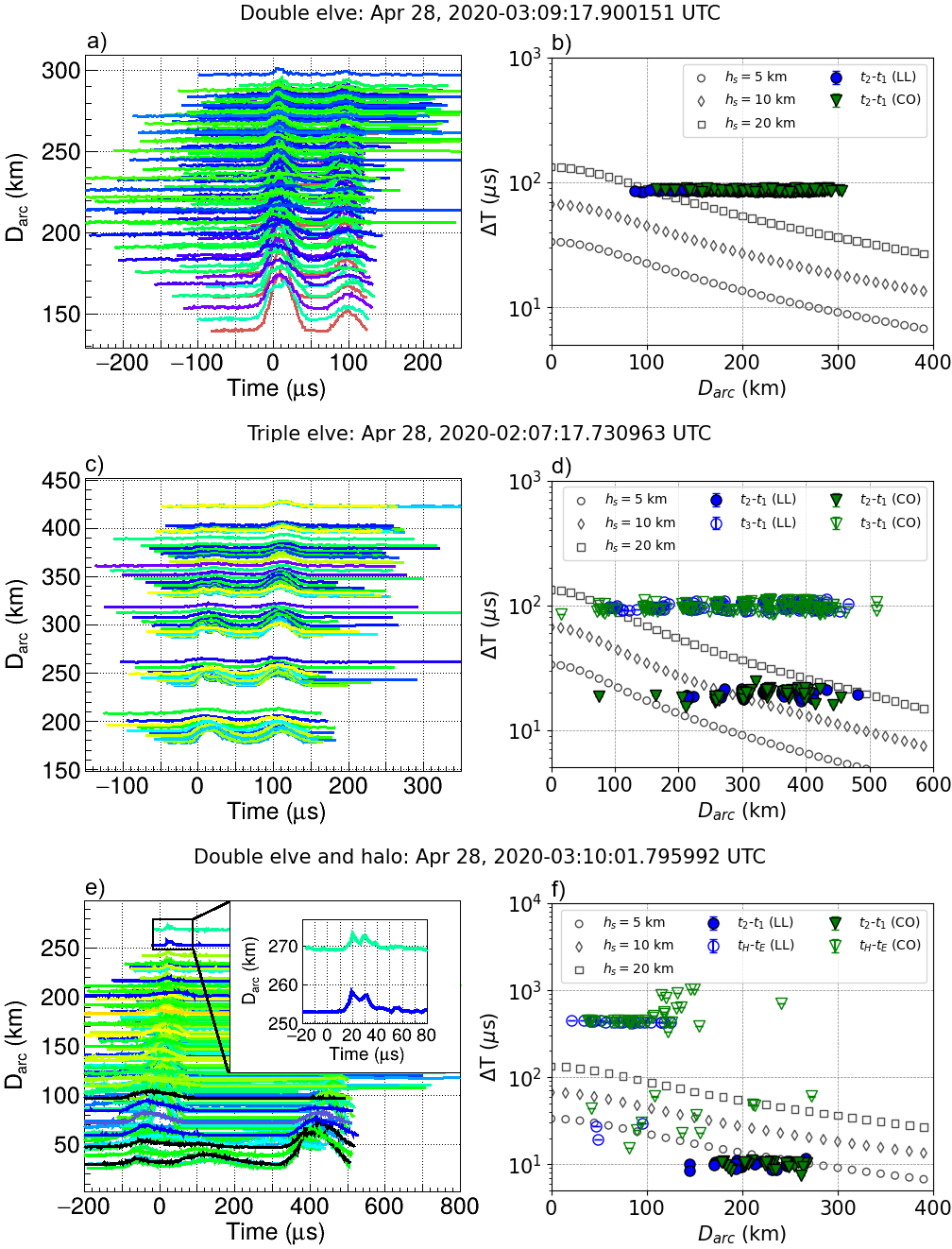}
    \caption{a) All traces of the double elve event from Figure \ref{fig:peakfinder} in Coihueco (CO), with the first peak aligned to the lightning source time ($t=0$). b) Time difference of the event at CO and Los Leones (LL), with a mean value of $(87\pm1) \,\mu$s. The bouncing mechanism model curves for 5, 10, and 20 km lightning heights are also displayed. c) Traces of the triple elve event from Figure \ref{fig:peakfinder} in CO, with the first peak aligned to the lightning source time ($t=0$). d) The result of $(t_2-t_1)$ and $(t_3-t_1)$ from LL and CO, with mean values of $(19 \pm 2)\mu$s and $(98 \pm 8) \,\mu$s respectively. e) Traces from an event showing multiple pulses that are significantly wider (halo) compared to typical elve pulses, occurring between 20 and 150 km of $\mathrm{D_{arc}}$. $t=0$ is the time of the lightning bolt occurring on April 28, 2020, at 03:10:01.795992 UTC. Beyond 150 km, traces of a double elve are observed (see zoomed-in plot). f) Time difference ($t_2 - t_1$) for the double elve, with a mean value of $(10 \pm 1)\, \mu$s. Other values represent the time difference between the double elve and halo traces ($t_H - t_E$). This event was detected at two FD buildings, CO and LL.}
    \label{fig:double_triple_halo}
\end{figure}

In panel a of Figure \ref{fig:double_triple_halo}, we present all traces of the double elve event from Figure \ref{fig:peakfinder}, aligning the first peak with the time of the lightning source. This plot shows that the time difference between the first and second peaks remains constant as the arc distance, $\mathrm{D_{arc}}$, increases. Panel b shows the calculation of $\Delta T_d = t_2 - t_1$ for these traces, corresponding to the event detected at Coihueco (CO), yielding a mean value of $\overline{\Delta T_d} = (87 \pm 1)\, \mu$s. This time difference is consistent across two detection buildings, as evidenced by the fact that $t_2 - t_1$ remains similar for traces of the same event detected at Los Leones (LL). These findings contradict the predictions of the ground reflection model proposed by \citeA{Marshall2015}, which is represented by curves decreasing with $\mathrm{D_{arc}}$ for three different lightning heights (5, 10, and 20 km) in the same plot.

A similar behaviour is observed in triple elve events, where $\Delta T_d = t_2 - t_1$ and $\Delta T_t = t_3 - t_1$ tend to remain constant. Panel c of Figure \ref{fig:double_triple_halo} displays the traces of the triple event discussed in Figure \ref{fig:peakfinder}. Notably, the time differences between peaks remain constant, as shown in the plot of $\Delta T$ vs. $\mathrm{D_{arc}}$ (panel d). In this case, we obtained average values of $ \overline{\Delta T_d} = (19 \pm 2) \, \mu$s and $ \overline{\Delta T_t} = (98 \pm 8)\, \mu$s at two FD buildings, Coihueco and Los Leones.

Upon analyzing the traces of events from the four storms listed in table \ref{tab:data}, we identified instances where $\overline{\Delta T_t} > 350\, \mu$s, which drew our attention. Upon further examination of the traces from these events, we observed that they involved a combination of two TLEs: elves and halos. Panel e of Figure \ref{fig:double_triple_halo} shows the traces of an event from the April 28, 2020 storm. In this example, traces between 20 and 150 km exhibit multiple pulses significantly wider than the typical elve pulses. Furthermore, beyond 150 km, we observe the traces of a double elve event (see zoomed-in plot) with an average $ \overline{\Delta T_d} = (10 \pm 1)\, \mu$s. In the $\Delta T$ vs. $\mathrm{D_{arc}}$ plot (panel f), we distinguish the $t_2 - t_1$ corresponding to the double elve event, while the other values represent the time difference between the elve and halo traces ($t_H - t_E$). This result was observed independently at Coihueco and Los Leones. The number of halos occurring with multi-elves events is listed in the table \ref{tab:data} with the tag 'Halo'.

\begin{figure}
    \centering
    \includegraphics[scale=0.5]{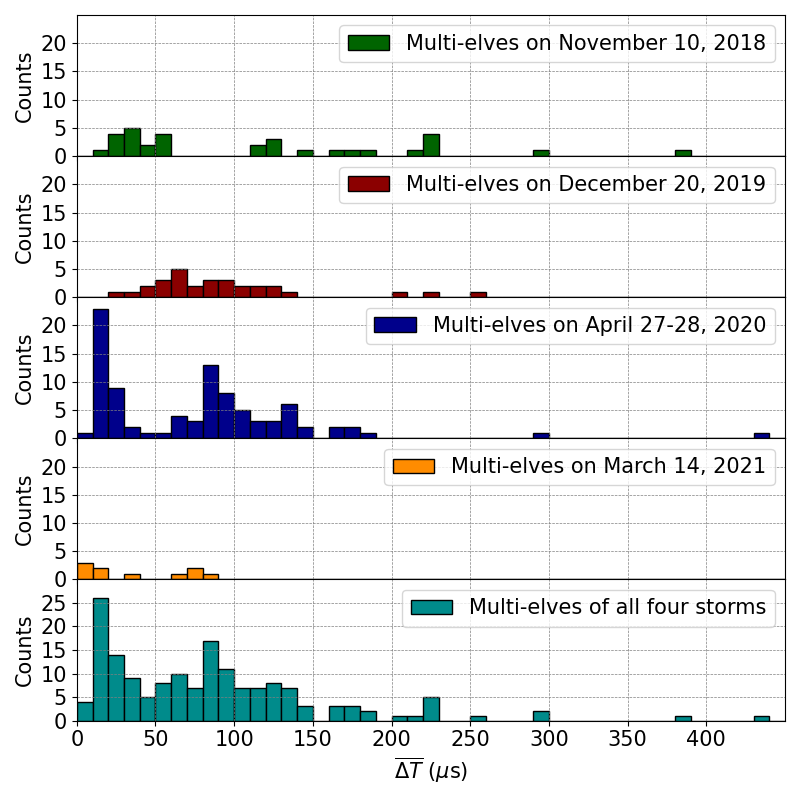}
    \caption{Distribution of the average time difference of the multi-elves of four storms detected by the FD. Each storm has a different distribution, with the April storm having the most events. The FD high temporal resolution allows reporting events with $\overline{\Delta T}< 10 \,\mu$s. The fifth histogram shows that most events fall within the 10 to 40 $\mu$s range and between 60 and 100 $\mu$s. Additionally, some events exhibit $\overline{\Delta T}$ values exceeding 200 $\mu$s, and in rare cases, around 450 $\mu$s, associated with combined elves and halo events.}
    \label{fig:dthisto}
\end{figure}

In Figure \ref{fig:dthisto}, we show the resulting histogram of the mean time delay of the events analyzed. Each storm exhibits a distinct distribution of $\overline{\Delta T}$, and it is evident that the storm of April 27-28, 2020, presents the highest occurrence of multi-elves. The high temporal resolution of the FD allows us to report events with $\Delta T$ as short as 7 $\mu$s. However, most events fall within 10 $\mu$s to 40 $\mu$s and 60 $\mu$s to 100 $\mu$s, as shown in the fifth histogram in the same figure. Furthermore, we have observed events with time delay exceeding 200 $\mu$s and, in some cases, around 450 $\mu$s. While these longer $\Delta T$ events may involve elves in conjunction with other TLEs such as halos, our current study does not delve into this aspect. Our primary focus remains on reporting the observed time delay of events exploiting the FD high temporal resolution.

\begin{figure}
    \centering
    \includegraphics[scale=0.5]{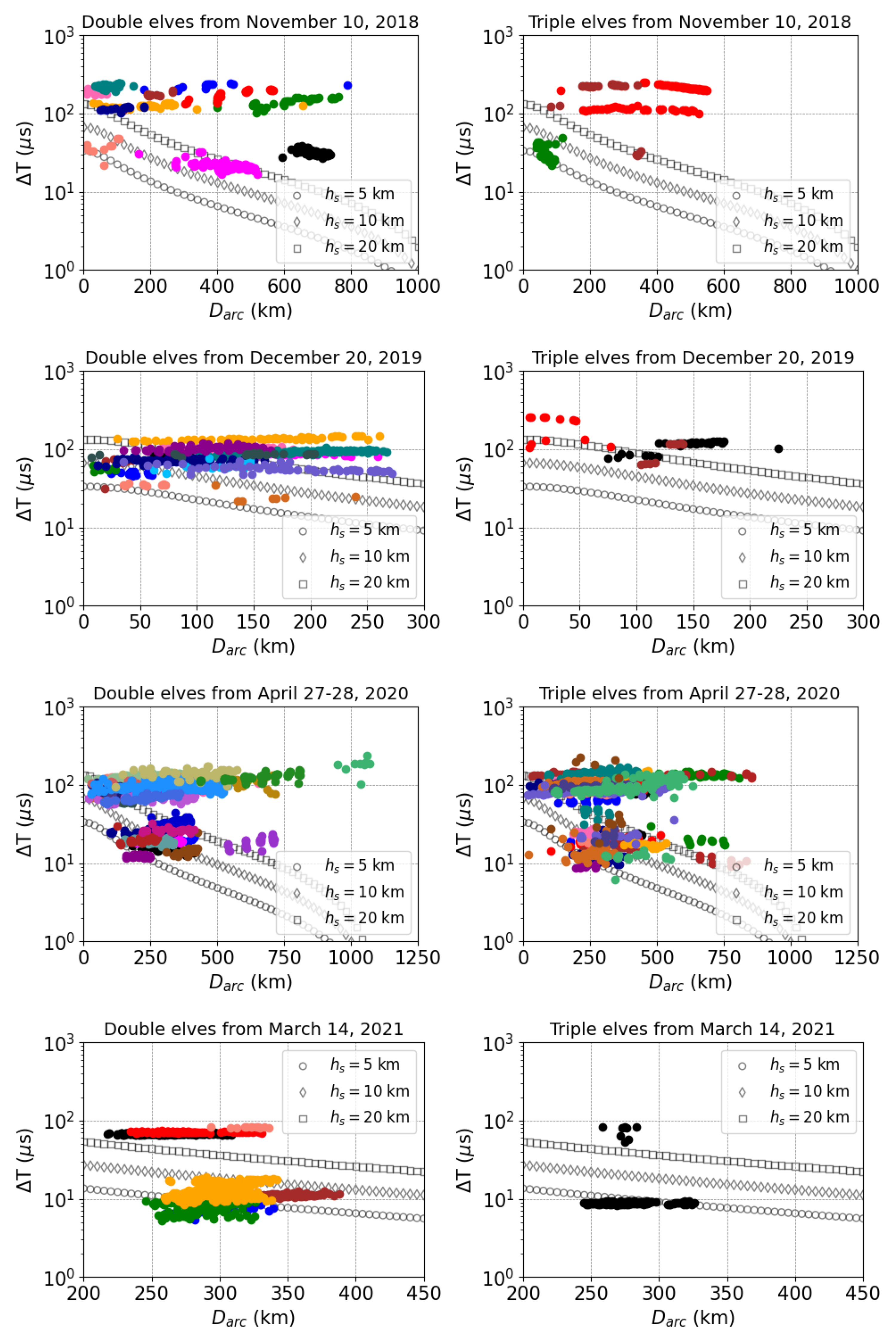}
    \caption{Temporal differences ($\Delta T$) for double (left) and triple (right) elves across the four selected storms. Each colour represents $\Delta T$ for a specific event. The empty markers illustrate various curves predicted by the model of \citeA{Marshall2015}, assuming intracloud lightning at height $h_\mathrm{s}$. The Auger FD data frequently show multi-elves with nearly constant $\Delta T$ values. This consistent $\Delta T$ trend is evident even in the March storm detected by HEAT at distances ranging from 200 to 400 km.}
    \label{fig:constantdt}
\end{figure}

Finally, Figure \ref{fig:constantdt} illustrates the temporal differences for double and triple elves across the four different storms, demonstrating that this result is independent of the specific characteristics of each storm. In these plots, each colour represents the $\Delta T$ of a particular event. On the left, we show the double elves, and on the right, the triple elves. Additionally, the curves with empty markers represent the reflection model, which does not fit any of the events.

\section{Discussion}
The temporal resolution of the FD cameras allowed us to capture a broad range of temporal difference values between the rings of the multi-elves. These events predominantly cluster into time delay values between 7 and 40 $\mu$s, and between 60 and 100 $\mu$s. This bimodal distribution may indicate distinct physical mechanisms underlying the generation of multi-elves.

To explore an alternative explanation for these observations, we adopted the time-domain model of the lightning electromagnetic pulse interaction with the lower ionosphere proposed by \citeA{marshall2012improved}, which suggests that double elves are a consequence of the rise time, $\tau_r$, and fall time, $\tau_f$, of the lightning pulse waveform.

We studied the waveforms from ENTLN sensors located in the region of northern Argentina within the field of view of the Auger observatory. The ENTLN~\cite{zhu2022upgrades} is a globally distributed ground-based network with 1,800 broadband electric sensors capable of detecting intracloud and cloud-to-ground flashes. On average, the ENTLN reports approximately 50 lightning events per second worldwide. These sensors record raw electric field waveforms within a frequency bandwidth of 1 Hz to 12 MHz, which are then processed using the time-of-arrival technique to determine real-time lightning geolocation. 

\begin{figure}[h]
    \centering
    \includegraphics[scale=0.45]{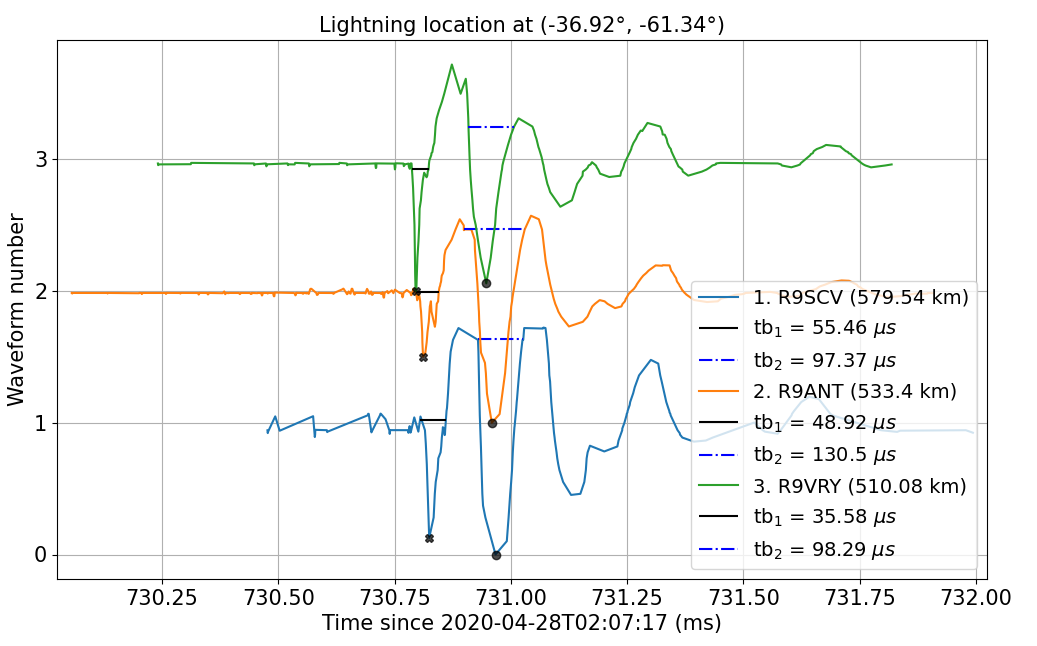}
    \caption{Waveforms from ENTLN sensors located within 500 and 600 km from the lightning correlated with the triple elve event shown in figures \ref{fig:peakfinder} and \ref{fig:double_triple_halo}. The most prominent peaks in each signal are identified using the \textit{find\_peaks} function from SciPy. The peak corresponding to the first component of the waveform is marked with a cross, while the peak corresponding to the second component is marked with a dot. The horizontal lines indicate the width of each peak at 95\% of the pulse amplitude, $t_{b1}$ and $t_{b2}$ respectively. The legend shows the sensor label and its distance from the source, with the corresponding $t_{b1}$ and $t_{b2}$ values.}
    \label{fig:waveform_triple}
\end{figure}

Figure \ref{fig:waveform_triple} shows the waveforms recorded by ENTLN sensors located 500 to 600 km from the lightning correlated with the triple elve presented in figures \ref{fig:peakfinder} and \ref{fig:double_triple_halo}. The most prominent peaks in each waveform are identified using SciPy's \textit{find\_peaks} function. The peak corresponding to the first waveform component is marked with a cross, and the second (skywave) is marked with a dot. Horizontal lines denote the width of each peak at 95\% of the pulse amplitude, defining the base times $t_{b1}$ and $t_{b2}$, respectively. In waveform analysis, the base time of pulses is typically measured at a percentage of their amplitude to mitigate overestimation. In this study, optimal values for the base time were determined at 95\% of the amplitude for 101 out of the 131 events correlated with the ENTLN. 

In these waveforms, we selected the skywave that reflects once in the ionosphere, as it is the most likely to generate the observed elves in the FD, exciting the molecules within it. Many of the signals correlated with elves exhibited saturation, despite the distance between the lightning strike and the antenna being between 500 and 600 km. Consequently, we conducted the study using the pulse base time ($t_{b2}$) rather than the individual rise and fall times, as it is challenging to determine $\tau_r$ and $\tau_f$. Furthermore, according to the model, $\tau_r$ and $\tau_f$ must be sufficiently large to separate the peaks of the multi-elves, implying that a longer $t_{b2}$ would result in a greater temporal difference ($\Delta T$) between these peaks.

\begin{figure}[h]
    \centering
    \includegraphics[scale=0.5]{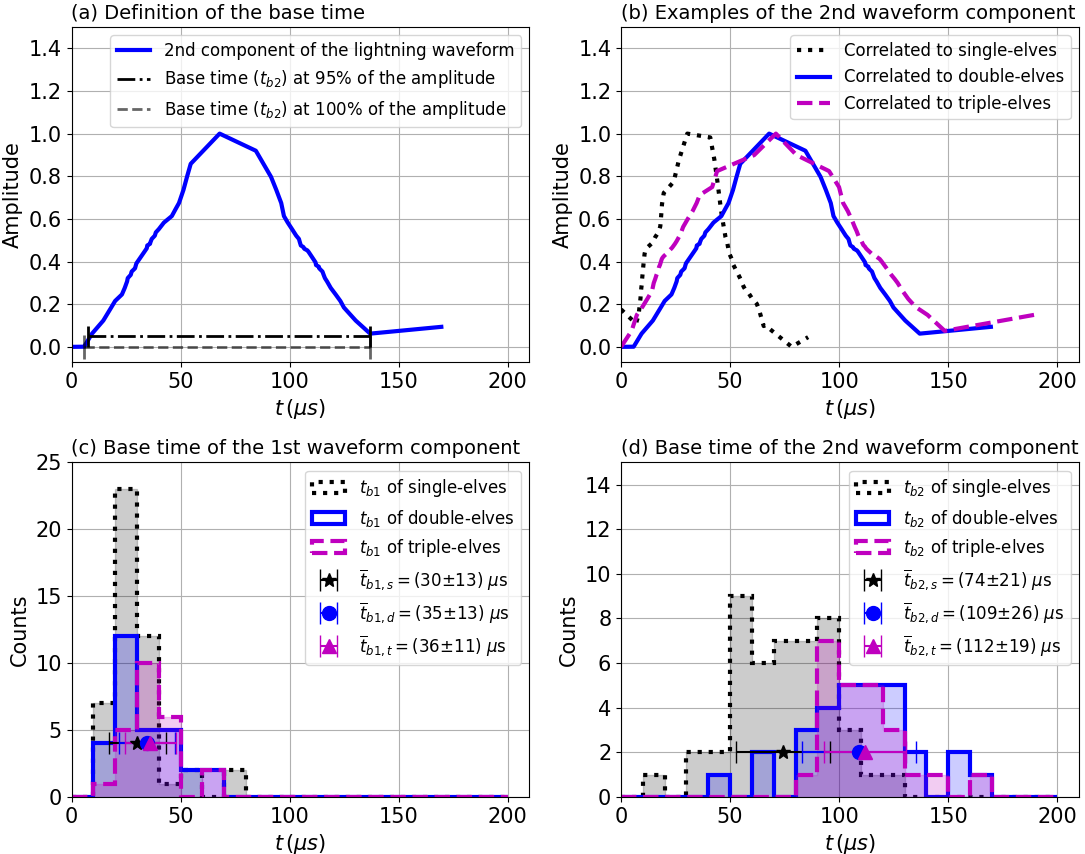}
    \caption{(a) Definition of the base time of the lightning waveform pulse given by Earth Networks. (b) Examples of the second component of lightning waveforms correlated with single (dotted line), double (solid line), and triple elves (dashed line). Distribution of the base time for the (c) first component and (d) second component of the waveform pulses correlated with elves detected on April 27-28, 2020. The mean value $\bar{t}_{b1}$ is comparable across events associated with single, double, and triple elves. In contrast, the mean value $\bar{t}_{b2}$ for single elves is $(74 \pm 21) \, \mu$s, which is significantly shorter than the mean values observed for double and triple elves, $(109 \pm 26) \, \mu$s and $(112 \pm 19) \, \mu$s, respectively.}
    \label{fig:waveforms}
\end{figure}

We observed that $t_{b2}$ associated with single elves tends to be narrower than those linked to multi-elves and that the pulse base time is correlated with the temporal difference between the rings of each event. An example of the pulse corresponding to the second component of a lightning waveform is shown in panel (a) of Figure \ref{fig:waveforms}. The dashed line indicates the base time of the pulse ($t_{b2}$) at 100\% of its amplitude, and the dashed-dotted line corresponds to 95\%. Panel (b) illustrates examples of a pulse correlated with a single elve event (dotted line), a double elve (solid line), and a triple elve (dashed line), with $t_{b2}$ values of 56, 130, and 148 $\mu$s, respectively.

In panels (c) and (d) of the same figure, the distribution of the base time of the first ($t_{b1}$) and second component of events correlated with single, double and triple elves is presented. It is noteworthy that, unlike the first component (which mean values are $\bar{t}_{b1} = (30 \pm 13) \,\mu s$, $(35 \pm 13) \,\mu s$ and $(36 \pm 11) \,\mu s$ for singles, doubles and triples, respectively), the base time of the second component is distinguishable between events correlated with single elves and multi-elves. The single elves distribution has a mean value of $\bar{t}_{b2} = (74 \pm 21) \,\mu s$, which is shorter than mean values from double and triple elves distributions, $(109 \pm 26) \,\mu s$ and $(112 \pm 19) \,\mu s$ respectively. These events correspond to the night of April 27-28, 2020, where we observed the highest proportion of multi-elves.

\begin{figure}[h]
    \centering
    \includegraphics[scale=0.5]{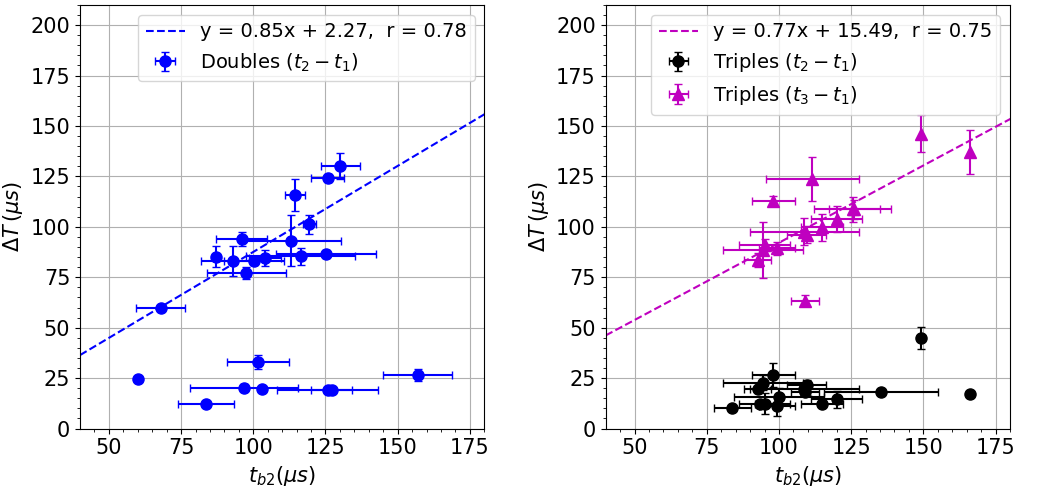}
    \caption{Base time of the second component of waveforms ($t_{b2}$) correlated with double elves (left) and triple elves (right) detected on April 27-28, 2020. In both cases, it can be observed that, beyond 50 $\mu$s of peak separation time for multi-elves ($\Delta T$), there is a high correlation with $t_{b2}$: $r=0.78$ in the fitting of doubles and, $r=0.75$ in the triple events. As the base time increases, the temporal gap between the peaks of multi-elves also widens.}
    \label{fig:tb_deltat}
\end{figure}

From the waveform of the same lightning event across multiple sensors, we calculated the average value and standard deviation of $t_{b2}$ for each event. Figure \ref{fig:tb_deltat} illustrates the relationship between $t_{b2}$ and the time gap between the rings of multi-elves. In this plot, it is evident that for values above 50 $\mu$s, $\Delta T$ increases with the base time of the pulse for both double and triple elve events. A strong correlation with $t_{b2}$ is observed, with a correlation coefficient of $r=0.78$ for double elves and $r=0.75$ for triple elves.

These results indicate that the origin of multiple rings of elves with $\Delta T \geq 50 \,\mu$s depends on the duration of the current density pulse of the lightning that produces them. As shown by \citeA{Liu2017}, short EIPs of 20 $\mu$s tend to produce double elves, whereas longer EIPs of 30 to 40 $\mu$s are more effective in separating the rings, resulting in the visualization of quadruplet elve events. It is important to note that these durations refer to the first component of the waveforms (known as the surface wave), while our analysis focuses on the second component, which is typically much longer.

Results below 50 $\mu$s do not demonstrate a clear relationship. Note that varying combinations of rise time and fall time can yield different $\Delta T$ values. In this context, using the base time, which is a combination of both, can result in the formation of these two subgroups in the graphs. However, these results align more closely with the predictions of \citeA{marshall2012improved} than with the bouncing mechanism \cite{Marshall2015}, because the time gap between the peaks of double and triple elves remains typically constant regardless of the arc distance to the lightning that generates them. We expected that events with $\Delta T < 100 \, \mu$s could be associated with intracloud lightning occurring at heights not exceeding 14 km. However, we observe that the decreasing curves predicted by the model do not align with the observations from the Auger FD even in the events with $\Delta T$ between 7 and 40 $\mu$s. These multi-elves with $\Delta T < 50\, \mu$s may originate from alternative mechanisms.

\section{Conclusions}\label{Conclusions}
We have found that the time gap between rings in both double and triple elves remains relatively constant with the arc distance to the lightning source. We frequently observed this behaviour in the Auger FD data and reported the time difference of multi-elves from four different storms occurring at various times and distances of the FD. Our study benefits from the high temporal resolution of the FD, which enables us to report double and triple ring events with temporal differences ranging from 7 $\mu$s to 260 $\mu$s and very rare events around 300 and 450 $\mu$s that are related to halos occurring with elves. 

These results suggest that the reflection of the EMP produced by intra-cloud lightning does not cause the multiple rings in the elves observed at the Pierre Auger Observatory. Contrary to the expected behaviour from the model, the temporal difference between the rings is nearly constant with the arc distance to the lightning source. 

On the other hand, multi-elves with a time gap exceeding 50 $\mu$s display a linear relationship with the base time of the second component of the lightning waveform, indicating that the generation of multiple rings is primarily governed by the duration of the lightning current density pulse rather than the bolt height. Although we report an unprecedented number of events with time delays shorter than 50 $\mu$s, we did not find a direct relationship with the pulse base time that could explain their origin. Moreover, within the temporal and spatial resolution of our measurements, none of the events matched the dependence of $\Delta T$ on $\mathrm{D_{arc}}$ predicted by the ground reflection model, which may suggest that the ground and atmospheric conditions in the field of view of the FD are not optimal for this mechanism or point to the involvement of a different underlying process. These results contribute to understanding the origin of multi-elves and provide a basis for further, more in-depth investigation into this phenomenon.

\section*{Open Research Section}
The analysis presented in this manuscript utilized the elves dataset available in \cite{vasquez_ramirez_2024_14224794}. 
These data files, in \texttt{.root} format, containing the traces of elves for each pixel triggered by the event in the Auger Observatory's fluorescence detector telescopes. The \texttt{multi-elves.ipynb} notebook includes the developed code used to analyze these traces. In addition, we used \texttt{ enipy-3}~\cite{stock2018enipy} to analyze the data of the lightning waveform. \texttt{enipy-3} is an open-source Python toolkit developed by Earth Networks under the MIT license and available at \url{https://bitbucket.org/earthnetworksrd/workspace/repositories/}. For further details on reproducing the figures in this manuscript, refer to the \texttt{README} file in \cite{vasquez_ramirez_2024_14224794}.

\acknowledgments
We acknowledge Jeff Lapierre of ENTLN for sharing the lightning waveform data, providing constructive suggestions, and engaging in discussions with us.

\begin{sloppypar}
The successful installation, commissioning, and operation of the Pierre
Auger Observatory would not have been possible without the strong
commitment and effort from the technical and administrative staff in
Malarg\"ue. We are very grateful to the following agencies and
organizations for financial support:
\end{sloppypar}

\begin{sloppypar}
Argentina -- Comisi\'on Nacional de Energ\'\i{}a At\'omica; Agencia Nacional de
Promoci\'on Cient\'\i{}fica y Tecnol\'ogica (ANPCyT); Consejo Nacional de
Investigaciones Cient\'\i{}ficas y T\'ecnicas (CONICET); Gobierno de la
Provincia de Mendoza; Municipalidad de Malarg\"ue; NDM Holdings and Valle
Las Le\~nas; in gratitude for their continuing cooperation over land
access; Australia -- the Australian Research Council; Belgium -- Fonds
de la Recherche Scientifique (FNRS); Research Foundation Flanders (FWO),
Marie Curie Action of the European Union Grant No.~101107047; Brazil --
Conselho Nacional de Desenvolvimento Cient\'\i{}fico e Tecnol\'ogico (CNPq);
Financiadora de Estudos e Projetos (FINEP); Funda\c{c}\~ao de Amparo \`a
Pesquisa do Estado de Rio de Janeiro (FAPERJ); S\~ao Paulo Research
Foundation (FAPESP) Grants No.~2019/10151-2, No.~2010/07359-6 and
No.~1999/05404-3; Minist\'erio da Ci\^encia, Tecnologia, Inova\c{c}\~oes e
Comunica\c{c}\~oes (MCTIC); Czech Republic -- GACR 24-13049S, CAS LQ100102401,
MEYS LM2023032, CZ.02.1.01/0.0/0.0/16{\textunderscore}013/0001402,
CZ.02.1.01/0.0/0.0/18{\textunderscore}046/0016010 and
CZ.02.1.01/0.0/0.0/17{\textunderscore}049/0008422 and CZ.02.01.01/00/22{\textunderscore}008/0004632;
France -- Centre de Calcul IN2P3/CNRS; Centre National de la Recherche
Scientifique (CNRS); Conseil R\'egional Ile-de-France; D\'epartement
Physique Nucl\'eaire et Corpusculaire (PNC-IN2P3/CNRS); D\'epartement
Sciences de l'Univers (SDU-INSU/CNRS); Institut Lagrange de Paris (ILP)
Grant No.~LABEX ANR-10-LABX-63 within the Investissements d'Avenir
Programme Grant No.~ANR-11-IDEX-0004-02; Germany -- Bundesministerium
f\"ur Bildung und Forschung (BMBF); Deutsche Forschungsgemeinschaft (DFG);
Finanzministerium Baden-W\"urttemberg; Helmholtz Alliance for
Astroparticle Physics (HAP); Helmholtz-Gemeinschaft Deutscher
Forschungszentren (HGF); Ministerium f\"ur Kultur und Wissenschaft des
Landes Nordrhein-Westfalen; Ministerium f\"ur Wissenschaft, Forschung und
Kunst des Landes Baden-W\"urttemberg; Italy -- Istituto Nazionale di
Fisica Nucleare (INFN); Istituto Nazionale di Astrofisica (INAF);
Ministero dell'Universit\`a e della Ricerca (MUR); CETEMPS Center of
Excellence; Ministero degli Affari Esteri (MAE), ICSC Centro Nazionale
di Ricerca in High Performance Computing, Big Data and Quantum
Computing, funded by European Union NextGenerationEU, reference code
CN{\textunderscore}00000013; M\'exico -- Consejo Nacional de Ciencia y Tecnolog\'\i{}a
(CONACYT) No.~167733; Universidad Nacional Aut\'onoma de M\'exico (UNAM);
PAPIIT DGAPA-UNAM; The Netherlands -- Ministry of Education, Culture and
Science; Netherlands Organisation for Scientific Research (NWO); Dutch
national e-infrastructure with the support of SURF Cooperative; Poland
-- Ministry of Education and Science, grants No.~DIR/WK/2018/11 and
2022/WK/12; National Science Centre, grants No.~2016/22/M/ST9/00198,
2016/23/B/ST9/01635, 2020/39/B/ST9/01398, and 2022/45/B/ST9/02163;
Portugal -- Portuguese national funds and FEDER funds within Programa
Operacional Factores de Competitividade through Funda\c{c}\~ao para a Ci\^encia
e a Tecnologia (COMPETE); Romania -- Ministry of Research, Innovation
and Digitization, CNCS-UEFISCDI, contract no.~30N/2023 under Romanian
National Core Program LAPLAS VII, grant no.~PN 23 21 01 02 and project
number PN-III-P1-1.1-TE-2021-0924/TE57/2022, within PNCDI III; Slovenia
-- Slovenian Research Agency, grants P1-0031, P1-0385, I0-0033, N1-0111;
Spain -- Ministerio de Ciencia e Innovaci\'on/Agencia Estatal de
Investigaci\'on (PID2019-105544GB-I00, PID2022-140510NB-I00 and
RYC2019-027017-I), Xunta de Galicia (CIGUS Network of Research Centers,
Consolidaci\'on 2021 GRC GI-2033, ED431C-2021/22 and ED431F-2022/15),
Junta de Andaluc\'\i{}a (SOMM17/6104/UGR and P18-FR-4314), and the European
Union (Marie Sklodowska-Curie 101065027 and ERDF); USA -- Department of
Energy, Contracts No.~DE-AC02-07CH11359, No.~DE-FR02-04ER41300,
No.~DE-FG02-99ER41107 and No.~DE-SC0011689; National Science Foundation,
Grant No.~0450696, and NSF-2013199; The Grainger Foundation; Marie
Curie-IRSES/EPLANET; European Particle Physics Latin American Network;
and UNESCO.
\end{sloppypar}

%
%

\bibliography{references}

@article{AbrahamEtal2010,
  title={The fluorescence detector of the {Pierre Auger Observatory}},
  author={Abraham, J. and Abreu, P. and Aglietta, M. and Aguirre, C. and Ahn, E.J. and Allard, D. and Allekotte, I. and Allen, J. and Allison, P. and Alvarez-Mu{\~n}iz, J. and others},
  journal={Nuclear Instruments and Methods in Physics Research Section A: Accelerators, Spectrometers, Detectors and Associated Equipment},
  volume={620},
  number={2-3},
  pages={227--251},
  year={2010},
  publisher={Elsevier}
}

@article{MussaCiaccio2012,
  title={Observation of ELVES at the {Pierre Auger Observatory}},
  author={Mussa, R and Ciaccio, G and Pierre Auger Collaboration and others},
  journal={The European Physical Journal Plus},
  volume={127},
  number={8},
  pages={94},
  year={2012},
  publisher={Springer}
}

@article{FukunishiEtal1996,
  title={Elves: Lightning-induced transient luminous events in the lower ionosphere},
  author={Fukunishi, H and Takahashi, Y and Kubota, M and Sakanoi, K and Inan, US and Lyons, WA},
  journal={Geophysical Research Letters},
  volume={23},
  number={16},
  pages={2157--2160},
  year={1996},
  publisher={Wiley Online Library}
}

@article{chen2008,
  title={Global distributions and occurrence rates of transient luminous events},
  author={Chen, Alfred B and Kuo, Cheng-Ling and Lee, Yi-Jen and Su, Han-Tzong and Hsu, Rue-Ron and Chern, Jyh-Long and Frey, Harald U and Mende, Stephen B and Takahashi, Yukihiro and Fukunishi, Hiroshi and others},
  journal={Journal of Geophysical Research: Space Physics},
  volume={113},
  number={A8},
  year={2008},
  publisher={Wiley Online Library}
}

@article{TaranenkoEtal1993,
  title={Interaction with the lower ionosphere of electromagnetic pulses from lightning: heating, attachment, and ionization},
  author={Taranenko, YN and Inan, US and Bell, TF},
  journal={Geophysical Research Letters},
  volume={20},
  number={15},
  pages={1539--1542},
  year={1993},
  publisher={Wiley Online Library}
}

@article{neubertEtal2020,
  title={A terrestrial gamma-ray flash and ionospheric ultraviolet emissions powered by lightning},
  author={Neubert, Torsten and {\O}stgaard, Nikolai and Reglero, Victor and Chanrion, Olivier and Heumesser, Matthias and Dimitriadou, Krystallia and Christiansen, Freddy and Budtz-J{\o}rgensen, Carl and Kuvvetli, Irfan and Rasmussen, Ib Lundgaard and others},
  journal={Science},
  volume={367},
  number={6474},
  pages={183--186},
  year={2020},
  publisher={American Association for the Advancement of Science}
}

@article{Merenda2020A,
  title={A Three Year Sample of Almost 1600 Elves Recorded Above {South America} by the {Pierre Auger} Cosmic-Ray {Observatory}},
  author={Aab, A and Abreu, P and Aglietta, M and Albuquerque, IFM and Albury, JM and Allekotte, I and others},
  journal={Earth and Space Science},
  pages={e2019EA000582},
  year={2020},
  publisher={Wiley Online Library}
}

@article{Newsome2010,
  title={Free-running ground-based photometric array imaging of transient luminous events},
  author={Newsome, RT and Inan, US},
  journal={Journal of Geophysical Research: Space Physics},
  volume={115},
  number={A7},
  year={2010},
  publisher={Wiley Online Library}
}

@article{Marshall2015,
  title={Elve doublets and compact intracloud discharges},
  author={Marshall, RA and Da Silva, CL and Pasko, Victor P},
  journal={Geophysical Research Letters},
  volume={42},
  number={14},
  pages={6112--6119},
  year={2015},
  publisher={Wiley Online Library}
}

@article{Lyu2015,
  title={Insights into high peak current in-cloud lightning events during thunderstorms},
  author={Lyu, Fanchao and Cummer, Steven A and McTague, Lindsay},
  journal={Geophysical Research Letters},
  volume={42},
  number={16},
  pages={6836--6843},
  year={2015},
  publisher={Wiley Online Library}
}

@article{Liu2017,
  title={Elves accompanying terrestrial gamma ray flashes},
  author={Liu, Ningyu and Dwyer, Joseph R and Cummer, Steven A},
  journal={Journal of Geophysical Research: Space Physics},
  volume={122},
  number={10},
  pages={10--563},
  year={2017},
  publisher={Wiley Online Library}
}

@article{marshall2012improved,
  title={An improved model of the lightning electromagnetic field interaction with the {D-region} ionosphere},
  author={Marshall, Robert A},
  journal={Journal of Geophysical Research: Space Physics},
  volume={117},
  number={A3},
  year={2012},
  publisher={Wiley Online Library}
}

@article{barrington1999elves,
  title={Elves triggered by positive and negative lightning discharges},
  author={Barrington-Leigh, Christopher P and Inan, Umran S},
  journal={Geophysical Research Letters},
  volume={26},
  number={6},
  pages={683--686},
  year={1999},
  publisher={Wiley Online Library}
}

@article{inan1991heating,
  title={Heating and ionization of the lower ionosphere by lightning},
  author={Inan, Umran S and Bell, Timothy F and Rodriguez, Juan V},
  journal={Geophysical Research Letters},
  volume={18},
  number={4},
  pages={705--708},
  year={1991},
  publisher={Wiley Online Library}
}

@article{capel2018mini,
  title={{Mini-EUSO}: A high resolution detector for the study of terrestrial and cosmic {UV} emission from the {International Space Station}},
  author={Capel, Francesca and Belov, Alexander and Casolino, Marco and Klimov, Pavel and JEM-EUSO Collaboration and others},
  journal={Advances in Space Research},
  volume={62},
  number={10},
  pages={2954--2965},
  year={2018},
  publisher={Elsevier}
}

@article{chern2003global,
  title={Global survey of upper atmospheric transient luminous events on the {ROCSAT-2} satellite},
  author={Chern, JL and Hsu, RR and Su, Han-Tzong and Mende, SB and Fukunishi, H and Takahashi, Y and Lee, Lou-Chuang},
  journal={Journal of Atmospheric and Solar-Terrestrial Physics},
  volume={65},
  number={5},
  pages={647--659},
  year={2003},
  publisher={Elsevier}
}

@article{witze2018inside,
  title={Inside {Argentina's} mega-storms},
  author={Witze, Alexandra},
  journal={Nature},
  volume={563},
  number={166},
  year={2018},
  publisher={Macmillan Publishers Ltd., London, England}
}

@article{MarshallEtal2008,
  title={Fast photometric imaging using orthogonal linear arrays},
  author={Marshall, Robert and Newsome, Robert and Inan, Umran},
  journal={IEEE transactions on geoscience and remote sensing},
  volume={46},
  number={11},
  pages={3885--3893},
  year={2008},
  publisher={IEEE}
}

@article{zhu2022upgrades,
  title={Upgrades of the {Earth Networks Total Lightning Network} in 2021},
  author={Zhu, Yanan and Stock, Michael and Lapierre, Jeff and DiGangi, Elizabeth},
  journal={Remote sensing},
  volume={14},
  number={9},
  pages={2209},
  year={2022},
  publisher={MDPI}
}

@article{pierre2015pierre,
  title={{The Pierre Auger cosmic ray observatory}},
  author={{The Pierre Auger Collaboration}},
  journal={Nuclear Instruments and Methods in Physics Research Section A: Accelerators, Spectrometers, Detectors and Associated Equipment},
  volume={798},
  pages={172--213},
  year={2015},
  publisher={Elsevier}
}

@inproceedings{stock2018enipy,
  title={Enipy, Python Tools for Working with Earth Networks Lightning Data},
  author={Stock, Michael and Lapierre, Jeff},
  booktitle={98th American Meteorological Society Annual Meeting},
  year={2018},
  organization={AMS}
}

@misc{vasquez_ramirez_2024_14224794,
doi          = {10.5281/zenodo.14224794},
url          = {https://doi.org/10.5281/zenodo.14224794},  
author       = {Vásquez-Ramírez, Adriana},
title        = {{Python Notebook and Dataset for Multi-Elves Analysis at the Pierre Auger Observatory}},
publisher    = {Zenodo},
year         = 2024,
type     = {dataset}
}

%
%
%
%
%
\newpage
\appendix
\section{The Pierre Auger Collaboration}
A.~Abdul Halim$^{13}$,
P.~Abreu$^{70}$,
M.~Aglietta$^{53,51}$,
I.~Allekotte$^{1}$,
K.~Almeida Cheminant$^{78,77}$,
A.~Almela$^{7,12}$,
R.~Aloisio$^{44,45}$,
J.~Alvarez-Mu\~niz$^{76}$,
A.~Ambrosone$^{44}$,
J.~Ammerman Yebra$^{76}$
% ...

\begin{description}[labelsep=0.2em,align=right,labelwidth=0.7em,labelindent=0em,leftmargin=2em,noitemsep,before={\renewcommand\makelabel[1]{##1 }}]
\item[$^{1}$] Centro At\'omico Bariloche and Instituto Balseiro (CNEA-UNCuyo-CONICET), San Carlos de Bariloche, Argentina
\item[$^{2}$] Departamento de F\'\i{}sica and Departamento de Ciencias de la Atm\'osfera y los Oc\'eanos, FCEyN, Universidad de Buenos Aires and CONICET, Buenos Aires, Argentina
\item[$^{3}$] IFLP, Universidad Nacional de La Plata and CONICET, La Plata, Argentina
\item[$^{4}$] Instituto de Astronom\'\i{}a y F\'\i{}sica del Espacio (IAFE, CONICET-UBA), Buenos Aires, Argentina
\item[$^{5}$] Instituto de F\'\i{}sica de Rosario (IFIR) -- CONICET/U.N.R.\ and Facultad de Ciencias Bioqu\'\i{}micas y Farmac\'euticas U.N.R., Rosario, Argentina
\item[$^{6}$] Instituto de Tecnolog\'\i{}as en Detecci\'on y Astropart\'\i{}culas (CNEA, CONICET, UNSAM), and Universidad Tecnol\'ogica Nacional -- Facultad Regional Mendoza (CONICET/CNEA), Mendoza, Argentina
\item[$^{7}$] Instituto de Tecnolog\'\i{}as en Detecci\'on y Astropart\'\i{}culas (CNEA, CONICET, UNSAM), Buenos Aires, Argentina
\item[$^{8}$] International Center of Advanced Studies and Instituto de Ciencias F\'\i{}sicas, ECyT-UNSAM and CONICET, Campus Miguelete -- San Mart\'\i{}n, Buenos Aires, Argentina
\item[$^{9}$] Laboratorio Atm\'osfera -- Departamento de Investigaciones en L\'aseres y sus Aplicaciones -- UNIDEF (CITEDEF-CONICET), Argentina
\item[$^{10}$] Observatorio Pierre Auger, Malarg\"ue, Argentina
\item[$^{11}$] Observatorio Pierre Auger and Comisi\'on Nacional de Energ\'\i{}a At\'omica, Malarg\"ue, Argentina
\item[$^{12}$] Universidad Tecnol\'ogica Nacional -- Facultad Regional Buenos Aires, Buenos Aires, Argentina
\item[$^{13}$] University of Adelaide, Adelaide, S.A., Australia
\item[$^{14}$] Universit\'e Libre de Bruxelles (ULB), Brussels, Belgium
\item[$^{15}$] Vrije Universiteit Brussels, Brussels, Belgium
\item[$^{16}$] Centro Brasileiro de Pesquisas Fisicas, Rio de Janeiro, RJ, Brazil
\item[$^{17}$] Centro Federal de Educa\c{c}\~ao Tecnol\'ogica Celso Suckow da Fonseca, Petropolis, Brazil
\item[$^{18}$] Instituto Federal de Educa\c{c}\~ao, Ci\^encia e Tecnologia do Rio de Janeiro (IFRJ), Brazil
\item[$^{19}$] Universidade de S\~ao Paulo, Escola de Engenharia de Lorena, Lorena, SP, Brazil
\item[$^{20}$] Universidade de S\~ao Paulo, Instituto de F\'\i{}sica de S\~ao Carlos, S\~ao Carlos, SP, Brazil
\item[$^{21}$] Universidade de S\~ao Paulo, Instituto de F\'\i{}sica, S\~ao Paulo, SP, Brazil
\item[$^{22}$] Universidade Estadual de Campinas (UNICAMP), IFGW, Campinas, SP, Brazil
\item[$^{23}$] Universidade Estadual de Feira de Santana, Feira de Santana, Brazil
\item[$^{24}$] Universidade Federal de Campina Grande, Centro de Ciencias e Tecnologia, Campina Grande, Brazil
\item[$^{25}$] Universidade Federal do ABC, Santo Andr\'e, SP, Brazil
\item[$^{26}$] Universidade Federal do Paran\'a, Setor Palotina, Palotina, Brazil
\item[$^{27}$] Universidade Federal do Rio de Janeiro, Instituto de F\'\i{}sica, Rio de Janeiro, RJ, Brazil
\item[$^{28}$] Universidad de Medell\'\i{}n, Medell\'\i{}n, Colombia
\item[$^{29}$] Universidad Industrial de Santander, Bucaramanga, Colombia
\item[$^{30}$] Charles University, Faculty of Mathematics and Physics, Institute of Particle and Nuclear Physics, Prague, Czech Republic
\item[$^{31}$] Institute of Physics of the Czech Academy of Sciences, Prague, Czech Republic
\item[$^{32}$] Palacky University, Olomouc, Czech Republic
\item[$^{33}$] CNRS/IN2P3, IJCLab, Universit\'e Paris-Saclay, Orsay, France
\item[$^{34}$] Laboratoire de Physique Nucl\'eaire et de Hautes Energies (LPNHE), Sorbonne Universit\'e, Universit\'e de Paris, CNRS-IN2P3, Paris, France
\item[$^{35}$] Univ.\ Grenoble Alpes, CNRS, Grenoble Institute of Engineering Univ.\ Grenoble Alpes, LPSC-IN2P3, 38000 Grenoble, France
\item[$^{36}$] Universit\'e Paris-Saclay, CNRS/IN2P3, IJCLab, Orsay, France
\item[$^{37}$] Bergische Universit\"at Wuppertal, Department of Physics, Wuppertal, Germany
\item[$^{38}$] Karlsruhe Institute of Technology (KIT), Institute for Experimental Particle Physics, Karlsruhe, Germany
\item[$^{39}$] Karlsruhe Institute of Technology (KIT), Institut f\"ur Prozessdatenverarbeitung und Elektronik, Karlsruhe, Germany
\item[$^{40}$] Karlsruhe Institute of Technology (KIT), Institute for Astroparticle Physics, Karlsruhe, Germany
\item[$^{41}$] RWTH Aachen University, III.\ Physikalisches Institut A, Aachen, Germany
\item[$^{42}$] Universit\"at Hamburg, II.\ Institut f\"ur Theoretische Physik, Hamburg, Germany
\item[$^{43}$] Universit\"at Siegen, Department Physik -- Experimentelle Teilchenphysik, Siegen, Germany
\item[$^{44}$] Gran Sasso Science Institute, L'Aquila, Italy
\item[$^{45}$] INFN Laboratori Nazionali del Gran Sasso, Assergi (L'Aquila), Italy
\item[$^{46}$] INFN, Sezione di Catania, Catania, Italy
\item[$^{47}$] INFN, Sezione di Lecce, Lecce, Italy
\item[$^{48}$] INFN, Sezione di Milano, Milano, Italy
\item[$^{49}$] INFN, Sezione di Napoli, Napoli, Italy
\item[$^{50}$] INFN, Sezione di Roma ``Tor Vergata'', Roma, Italy
\item[$^{51}$] INFN, Sezione di Torino, Torino, Italy
\item[$^{52}$] Istituto di Astrofisica Spaziale e Fisica Cosmica di Palermo (INAF), Palermo, Italy
\item[$^{53}$] Osservatorio Astrofisico di Torino (INAF), Torino, Italy
\item[$^{54}$] Politecnico di Milano, Dipartimento di Scienze e Tecnologie Aerospaziali , Milano, Italy
\item[$^{55}$] Universit\`a del Salento, Dipartimento di Matematica e Fisica ``E.\ De Giorgi'', Lecce, Italy
\item[$^{56}$] Universit\`a dell'Aquila, Dipartimento di Scienze Fisiche e Chimiche, L'Aquila, Italy
\item[$^{57}$] Universit\`a di Catania, Dipartimento di Fisica e Astronomia ``Ettore Majorana``, Catania, Italy
\item[$^{58}$] Universit\`a di Milano, Dipartimento di Fisica, Milano, Italy
\item[$^{59}$] Universit\`a di Napoli ``Federico II'', Dipartimento di Fisica ``Ettore Pancini'', Napoli, Italy
\item[$^{60}$] Universit\`a di Palermo, Dipartimento di Fisica e Chimica ''E.\ Segr\`e'', Palermo, Italy
\item[$^{61}$] Universit\`a di Roma ``Tor Vergata'', Dipartimento di Fisica, Roma, Italy
\item[$^{62}$] Universit\`a Torino, Dipartimento di Fisica, Torino, Italy
\item[$^{63}$] Benem\'erita Universidad Aut\'onoma de Puebla, Puebla, M\'exico
\item[$^{64}$] Unidad Profesional Interdisciplinaria en Ingenier\'\i{}a y Tecnolog\'\i{}as Avanzadas del Instituto Polit\'ecnico Nacional (UPIITA-IPN), M\'exico, D.F., M\'exico
\item[$^{65}$] Universidad Aut\'onoma de Chiapas, Tuxtla Guti\'errez, Chiapas, M\'exico
\item[$^{66}$] Universidad Michoacana de San Nicol\'as de Hidalgo, Morelia, Michoac\'an, M\'exico
\item[$^{67}$] Universidad Nacional Aut\'onoma de M\'exico, M\'exico, D.F., M\'exico
\item[$^{68}$] Institute of Nuclear Physics PAN, Krakow, Poland
\item[$^{69}$] University of \L{}\'od\'z, Faculty of High-Energy Astrophysics,\L{}\'od\'z, Poland
\item[$^{70}$] Laborat\'orio de Instrumenta\c{c}\~ao e F\'\i{}sica Experimental de Part\'\i{}culas -- LIP and Instituto Superior T\'ecnico -- IST, Universidade de Lisboa -- UL, Lisboa, Portugal
\item[$^{71}$] ``Horia Hulubei'' National Institute for Physics and Nuclear Engineering, Bucharest-Magurele, Romania
\item[$^{72}$] Institute of Space Science, Bucharest-Magurele, Romania
\item[$^{73}$] Center for Astrophysics and Cosmology (CAC), University of Nova Gorica, Nova Gorica, Slovenia
\item[$^{74}$] Experimental Particle Physics Department, J.\ Stefan Institute, Ljubljana, Slovenia
\item[$^{75}$] Universidad de Granada and C.A.F.P.E., Granada, Spain
\item[$^{76}$] Instituto Galego de F\'\i{}sica de Altas Enerx\'\i{}as (IGFAE), Universidade de Santiago de Compostela, Santiago de Compostela, Spain
\item[$^{77}$] IMAPP, Radboud University Nijmegen, Nijmegen, The Netherlands
\item[$^{78}$] Nationaal Instituut voor Kernfysica en Hoge Energie Fysica (NIKHEF), Science Park, Amsterdam, The Netherlands
\item[$^{79}$] Stichting Astronomisch Onderzoek in Nederland (ASTRON), Dwingeloo, The Netherlands
\item[$^{80}$] Universiteit van Amsterdam, Faculty of Science, Amsterdam, The Netherlands
\item[$^{81}$] Case Western Reserve University, Cleveland, OH, USA
\item[$^{82}$] Colorado School of Mines, Golden, CO, USA
\item[$^{83}$] Department of Physics and Astronomy, Lehman College, City University of New York, Bronx, NY, USA
\item[$^{84}$] Michigan Technological University, Houghton, MI, USA
\item[$^{85}$] New York University, New York, NY, USA
\item[$^{86}$] University of Chicago, Enrico Fermi Institute, Chicago, IL, USA
\item[$^{87}$] University of Delaware, Department of Physics and Astronomy, Bartol Research Institute, Newark, DE, USA
\item[] -----
\item[$^{a}$] Max-Planck-Institut f\"ur Radioastronomie, Bonn, Germany
\item[$^{b}$] also at Kapteyn Institute, University of Groningen, Groningen, The Netherlands
\item[$^{c}$] School of Physics and Astronomy, University of Leeds, Leeds, United Kingdom
\item[$^{d}$] Fermi National Accelerator Laboratory, Fermilab, Batavia, IL, USA
\item[$^{e}$] Pennsylvania State University, University Park, PA, USA
\item[$^{f}$] Colorado State University, Fort Collins, CO, USA
\item[$^{g}$] Louisiana State University, Baton Rouge, LA, USA
\item[$^{h}$] now at Graduate School of Science, Osaka Metropolitan University, Osaka, Japan
\item[$^{i}$] Institut universitaire de France (IUF), France
\item[$^{j}$] now at Technische Universit\"at Dortmund and Ruhr-Universit\"at Bochum, Dortmund and Bochum, Germany
\end{description}

\end{document}